\documentclass[onecolumn,showpacs,amsmath,amssymb]{revtex4-1}
\usepackage{abbreviations}
\usepackage{epsfig}
\begin{document}
\thispagestyle{empty}

\title{On the Casimir entropy for a ball in front of a plane}

\author{M. Bordag\footnote{bordag@itp.uni-leipzig.de}}
\affiliation{ Leipzig University,
Institute for Theoretical Physics,  04109 Leipzig, Germany}

\author{I.G. Pirozhenko\footnote{pirozhen@theor.jinr.ru}}
\affiliation{ Bogoliubov Laboratory of Theoretical Physics,\\
\small Joint Institute for Nuclear Research and Dubna International University\\
\small Dubna 141 980,
Russia}
\date{\small October 2, 2010}

\begin{abstract}
The violation of the third law of thermodynamics for metals described by the Drude model and for dielectrics with finite \DC conductivity is one of the most interesting problems in the field of the Casimir effect. It manifests itself as a non-vanishing of the entropy for vanishing temperature. We  review the relevant calculations for plane surfaces and calculate the corresponding contributions for a ball in front of a plane. In this geometry, these appear  in much the same way as for parallel planes. We conclude that the violation of the 3rd law is not related to the infinite size of the planes.
\end{abstract}
\maketitle
%%%%%%%%%%%%%%%%%%%%%%%%%%%%%%%%%%%%%%%%%%%%%%%%%%%%%%%%%%%%%%%%%%
\section{Introduction}
The violation of the third law of thermodynamics by the Casimir effect for certain properties of the interacting bodies is still one of the most interesting problems in the field. It manifests itself as a non vanishing limit of the entropy $S$,  defined as minus the derivative of the free energy $\F$ with respect to the temperature $T$,
\be
\label{1S}   \S=-\frac{\pa \F}{\pa T}\,,
\ee
for $T\to0$. It was first observed in \cite{beze02-66-062112,beze04-69-022119} for metals described by the Drude model and in \cite{geye05-72-085009} for dielectrics with \DC conductivity.
Inserting the corresponding permittivities into the Lifshitz formula for the free energy $\F$, at vanishing temperature a term linear in $T$ remains which by means of \Ref{1S} results in a non vanishing contribution to $\S$ at $T=0$. Obviously, the use of these permittivities, which otherwise work fine, when inserted into the Lifshitz formula, results in a behavior which is not only   non acceptable from a principle point of view but which is also in quite clear disagreement with experimental observations. Details  on this topic can be found in the recent review \cite{klim09-81-1827} (and also in the book \cite{BKMM}).

The Drude model is characterized by a permittivity
\be
\label{1D}\ep^{\rm D}(i \xi)=1+\frac{\om_p^2}{\xi(\ga+\xi)}\,,
\ee
where $\om=i\xi$ is the imaginary frequency, $\om_p$ is the plasma frequency and $\ga$ is the relaxation parameter. For $\ga=0$, the permittivity $\ep^{\rm D}(i\xi)$ turns into that of the plasma model which does not cause  problems with thermodynamics. A dielectric with \DC conductivity is characterized by a permittivity
\be
\label{1dc}\ep^{\rm dc}(i \xi)=\ep_0(i\xi)+\frac{4\pi \sigma_0}{\xi}\,,
\ee
where $\sigma_0$ is the \DC conductivity and $\ep_0(i\xi)$ is the permittivity of a dielectric without \DC conductivity of which we only need to know that it takes a finite limit $\ep_0\equiv\ep_0(0)$ for zero frequency.

In both cases the violation occurs if the parameters $\sigma_0$ or $\ga$ are non zero, depend on the temperature and decrease for $T\to0$. This happens for some reasonable idealizations of real materials where  $\sigma_0$ decreases exponentially fast or $\ga$ as a power of $T$ (for metals with perfect crystal lattice).

In the present paper we  reconsider the derivation of the violation terms and establish that an arbitrary slow decrease already results in a violation of the third law. For this purpose, we employ a representation different from that used in \cite{BKMM} which looks technically more direct and which allows for a deeper insight into the corresponding structures. We would like to stress that we make no claim about the physical reality of slowly decreasing parameters $\ga$ and $\sigma$.

During the past decade there was quite a number of attempts to avoid a violation of the third law. Most of them point to a modification by including additional physical effects.  An example is the addition of impurities to a perfect crystal lattice \cite{bost04-339-53}. Other consist in using impedance boundary conditions in place of the Drude model in the Lifshitz formula \cite{beze02-65-012111,esqu03-68-052103,brev05-71-056101,elli09-161-012010}. It must be admitted that no satisfactory understanding was reached so far.

In the present paper we answer the question whether a finite size of one of the interacting bodies is able to prevent the violation. For this we consider a sphere in front of a plane at low temperature. This is an extension of our previous paper \cite{bord10-1007}. We consider a sphere with the permittivities \Ref{1D} and \Ref{1dc} in front of a conducting plane. Special attention is paid to the case of small separation. We mention that the configuration of a ball in front of a plane at finite temperature is under active discussion, see for example \cite{Cana10-104-040403} and \cite{gies10-25-2279}. 

In the next section we reconsider the case of parallel planes and in the third section we treat the sphere-plane configuration. Conclusions are drawn in the last section. \\
Throughout the paper we use units with $\hbar =c=1$.

%%%%%%%%%%%%%%%%%%%%%%%%%%%%%%%%%%%%%%%%%%%%%%%%%%%%%%%%%%%%%%%%%%%%%%%%%%%%%
\section{The free energy for parallel planes}
The free energy for two parallel plane bodies at separation $a$ is given by the Lifshitz formula
\be \F=\frac{T}{2}\sum_{l=-\infty}^{\infty}
    \int\frac{d^2k}{(2\pi)^2}\sum_{i=\TE,\TM}\ln\left(1-r_{i}^2\,e^{-2aq}\right)
\label{2F}
\ee
with
\be q=\sqrt{\xi_l^2+k^2}\,,
\label{2q}
\ee
the Matsubara frequencies
\be \xi_l=2\pi l T\,,
\label{2MF}
\ee
and $\vec{k}$ is the momentum parallel to the planes. The reflection coefficients are
\be r_{\TE}=\frac{q-\sqrt{(\ep-1)\xi_l^2+q^2}}{q+\sqrt{(\ep-1)\xi_l^2+q^2}},
\label{2rTE}
\ee
for the TE mode and
\be r_{\TM}=\frac{\ep q-\sqrt{(\ep-1)\xi_l^2+q^2}}{\ep q+\sqrt{(\ep-1)\xi_l^2+q^2}},
\label{2rTM}
\ee
for the TM mode, where for $\ep$ one must insert one of the two, \Ref{1D} or \Ref{1dc},
according to the model considered.

Using the Abel-Plana formula, representation \Ref{2F} can be rewritten as a sum,
\be \F=E_0+\Delta_{\rm T}\F,
\label{2F1}
\ee
of the vacuum energy,
\be E_0=\frac12\int_{0}^{\infty}\frac{d\xi}{\pi}
     \int\frac{d^2k}{(2\pi)^2}\sum_{i=\TE,\TM}\ln\left(1-r_{i}^2\,e^{-2aq}\right),
\label{2E0}
\ee
depending on $T$ only through $\ga$ or $\sigma$, and the explicitely temperature dependent part of the free energy,
\be \Delta_{\rm T}\F=\frac{1}{4\pi^2}\int_0^\infty dx\,n_T(x)\ \Phi(x),
\label{2dF}
\ee
where
\be n_T(x)=\frac{1}{e^{x/T}-1}
\label{2BF}
\ee
is the Boltzmann factor. In \Ref{2dF} the notation
\be \Phi(x)=i\left(\varphi(i x)-\varphi(-ix)\right)
\label{2Fi}
\ee
is used which contains the two contributions
 appearing from turning the integration path $\xi\to\pm i x$ in the Abel-Plana formula and
\be \varphi(\xi)=\int_0^\infty dk\, k\ \sum_{i=\TE,\TM}\ln\left(1-r_{i}^2\,e^{-2aq}\right),
\label{2fi}
\ee
is the function which must be analytically continued. This representation of the free energy
is well known and it is especially useful at low temperature where the
Matsubara sum in \Ref{2F} converges slowly.

It must be mentioned that the division in \Ref{2F1} is done according to the contributions of the electromagnetic excitations. The vacuum energy $E_0$ does not contain contributions from the thermal photons whereas $\Delta_T\F$ is just their contribution. However, in the case of temperature dependent permittivity, the thermal excitation of the degrees of freedom of the interacting bodies like electronic or phonon excitations enter the vacuum energy through the corresponding parameters $\ga(T)$ or $\sigma(T)$. In this way their vacuum energy contributes also to the entropy which thus consists of two parts,
\be \S=\S_0+\S_1
\label{2S}
\ee
with
\be \S_0=-\frac{\pa E_0}{\pa T}=-\frac{\pa E_0}{\pa \mu}\,\frac{\pa \mu}{\pa T},
\label{2S0}
\ee
where $\mu$ is one of the two, $\ga$ or $\sigma$, and
\be \S_1=-\frac{\pa \Delta_T\F}{\pa T}.
\label{2S1}
\ee
The first part, $\S_0$ depends on the thermal excitations of the interacting bodies only, whereas the second part, $\S_1$, depends on the thermal excitations of the photons too.

The more conventional approach to derive the  violation terms,
used in the literature (see the book \cite{BKMM} for a detailed representation),
rests on the observation that the interesting terms result from
the $(l=0)$-contribution to the Matsubara sum in \Ref{2F}. As it
turned out,
 for metallic bodies described by the
Drude model the linear term is
\be     \F^{\rm Drude}=
        -\F^{\rm plasma, TE}_{l=0}+\dots\,
        =-\frac{T}{16\pi a^2}\,f^{\rm D}_{\rm lin}+\dots\,
\label{2FD} \ee
with
\be        f^{\rm D}_{\rm lin}=
      (2a)^2  \int_0^\infty dq\, q\,
        \ln\left(1-\left(r_{\rm plasma}^{\rm TE}\right)^2\ e^{-2aq}\right)\,.
\label{2DLin}\ee
Here $r_{\rm plasma}^{\rm TE}$ is the reflection coefficient of the TE
mode with the permittivity of the plasma model, i.e., $\ep^{\rm
D}(i\xi)$, \Ref{1D}, with $\ga=0$. For large $\om_p$ the linear in
$T$ contribution  takes the limiting value
\be  {\F^{\rm Drude}}_{|_{\om_p\to\infty}}=\frac{T}{16\pi a^2}\,
\zeta(3)+\dots\,. \label{2Fp} \ee
For a dielectric body with \DC conductivity, the contribution
linear in $T$ can be written as a difference between the
$(l=0)$-contributions of the TM mode with \DC conductivity and
without ($\sigma_0=0$ in Eq.\Ref{1dc}),
\be \F^{\rm dc}=
        \F^{\rm dc, TM}_{l=0}-\F^{\rm no~ dc, TM}_{l=0}+\dots\,\
        =-\frac{T}{16\pi a^2}   \, f^{\rm DC}_{\rm lin}+\dots\,
\label{2Fdc}
\ee
with
\be     f^{\rm DC}_{\rm lin}=
        \zeta(3)-{\rm Li}_3(r_0^2)\,
\label{2fDlin}\ee
and the notation
\be r_0=\frac{\ep_0-1}{\ep_0+1}\label{2r0}\ee
for  the TM reflection
coefficient for static permittivity. ${\rm Li}_3$ is a
polylogarithm function.

It is interesting to remember that the $(l=0)$-contribution to the Matsubara sum gives just the leading order contribution for $T\to\infty$. In this way, the violating terms are closely related to the high temperature limit, i.e., to the classical limit. It is, however, not clear whether this has any deeper meaning.

In the following subsections we use representation \Ref{2F1} with \Ref{2E0} and \Ref{2dF} and re-calculate the low temperature behavior in both models, Drude and \DC conductivity.
%%%%%%%%%%%%%%%%%%%%%%%%%%%%%%%%%%%%%%%%%%%%%%%%%%%%%%%%%%%%%%%%%%%%%%%%%%%%%%%%%%%%%%%%%%%%
\subsection{The vacuum energy in the Drude model}
In this subsection we calculate the contribution of the vacuum energy, $E_0(\ga)$, to the entropy at small temperature which comes in through the temperature dependence of the relaxation parameter $\ga(T)$. According to \Ref{2S0} we have
\be S_0=-\frac{\pa E_0(\ga)}{\pa \ga}\,\frac{\pa \ga}{\pa T}\,.
\label{21S0}
\ee
We assume a decrease of $\ga(T)$ for $T\to0$ according to
\be \ga(T)=\ga_1 T^\al+\dots
\label{21Tal}\ee
with $\al>0$. Therefore we need the expansion of the vacuum energy for small $\ga$. As it follows from the calculations below, this expansion contains a logarithmic contribution,
\be E_0(\ga)=E_0(0)+\ga\left(-\ln(2a\ga)\,\tilde{E}_1+E_1\right)+\dots\,.
\label{21E0}
\ee
Here, $E_0(0)$ is the vacuum energy of the plasma model and it does not contribute to $S_0$. The logarithmic term results from the TE mode. With \Ref{2E0}, \Ref{2rTE} and \Ref{1D} we note
\be E_0^{\rm TE}(\ga)=\frac{1}{4\pi^2}\int_0^\infty dq \,q\,\int_0^q d\xi\
                        h_q\left(\frac{\ga}{\xi}\right)\,,
\label{21E1}
\ee
where we defined
\be h_q(z)=\ln\left[ 1-
\left(\frac{q-\sqrt{\frac{\om_p^2}{1+z}+q^2}}{q+\sqrt{\frac{\om_p^2}{1+z}+q^2}}\right)^2
\, e^{-2aq}\right] \,.
\label{21fq}
\ee
As compared with \Ref{2E0} we changed the integration over $k$ for $q=\sqrt{\xi^2+k^2}$ and interchanged the orders of the integrations.

\begin{figure}[b]
\hspace{1cm}\includegraphics[width=7cm]{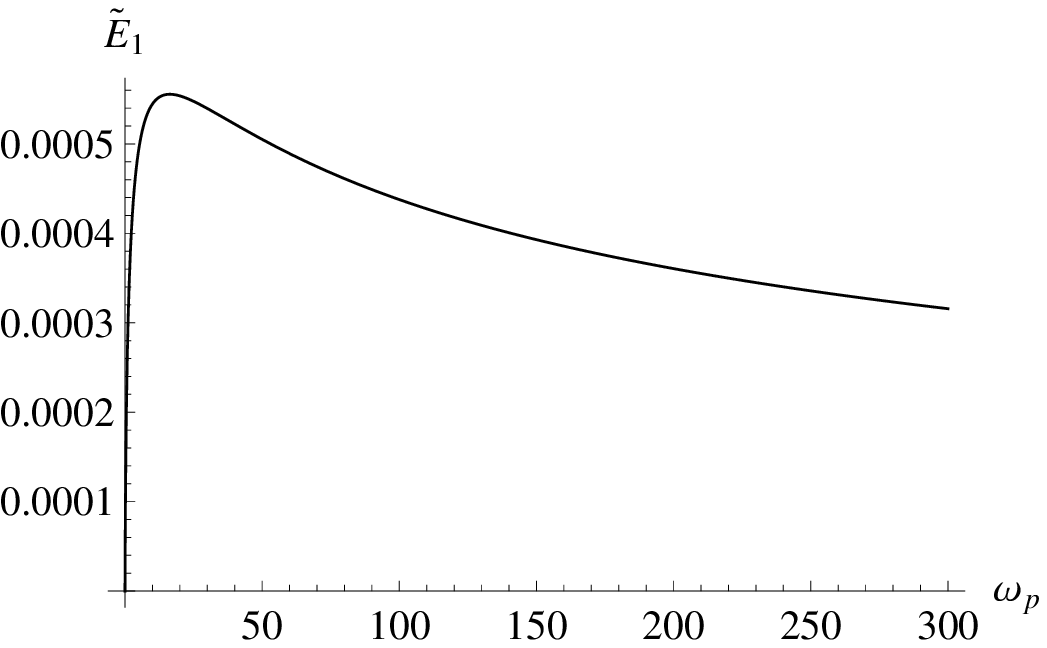}    \includegraphics[width=7cm]{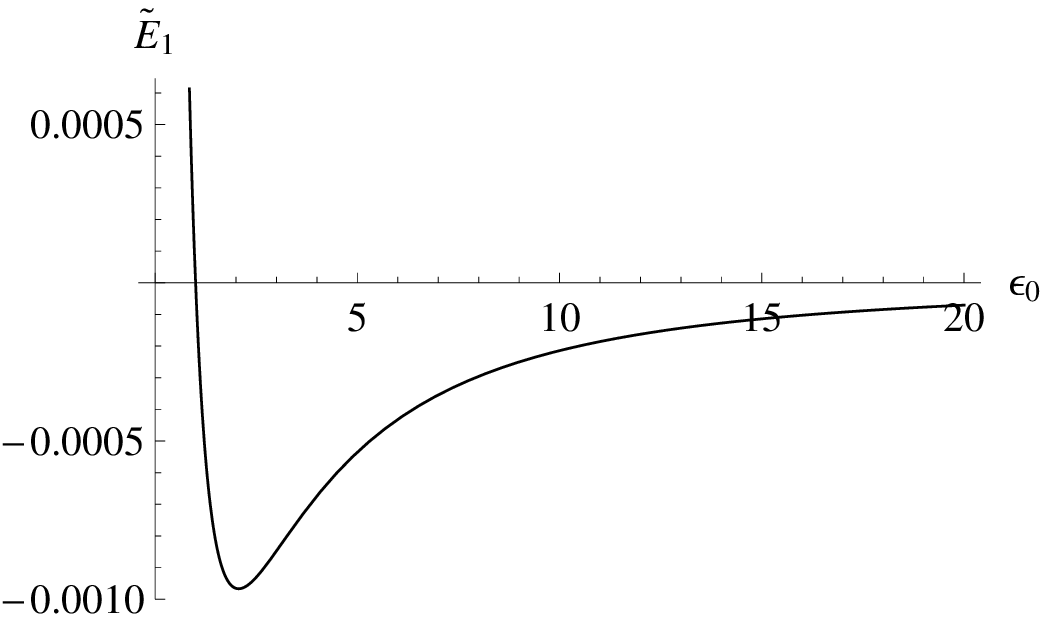}
\caption{The coefficient $\tilde{E}_1$ in \Ref{21E0} for the Drude model, \Ref{21Et}, as a function of $\om_p$ for a separation $a=1$ (left panel) and for a dielectric  with \DC conductivity, \Ref{22Et}, as a function of $\ep_0$ (right panel).}
\label{figDrudeTE}\label{figDCTM}
\end{figure}

In \Ref{21E1}, a formal expansion for small $\ga$ goes in powers of $\ga/\xi$ and already in the first order the $\xi$-integration becomes singular. For this reason we change the variable $\xi$ for $z=\ga/\xi$,
\be E_0^{\rm TE}(\ga)=\frac{1}{4\pi^2}\int_0^\infty dq \,q\,\int_{\ga/q}^\infty dz \,
                  \frac{  \ga}{z^2}  \,  h_q\left(z\right)\,,
\label{21E2}
\ee
and integrate by parts two times,
\be E_0^{\rm TE}(\ga)=\frac{1}{4\pi^2}\int_0^\infty dq \,q\,
\left\{ q h_q\left(\frac{\ga}{q}\right)  -\ga\ln\left(\frac{\ga}{q}\right)h_q'\left(\frac{\ga}{q}\right)
-\ga\int_{\ga/q}^\infty dz\, \ln z\,h_q^{''}(z)   \right\}\,.
\label{21E3}
\ee
Here the primes denote  derivatives of the function $h_q(z)$ with respect to $z$ and
we used the property of this function and of its derivatives to decrease sufficiently fast for $z\to\infty$. In representation \Ref{21E3}, it is possible to expand up to the first order in $\ga$ and we get for the expansion parameters in \Ref{21E0}
\be \tilde{E}_1=\frac{1}{4\pi^2}\int_0^\infty dq \,q\,h'_q(0)
\label{21Et}
\ee
and
\be E_1^{\rm TE}=\frac{1}{4\pi^2}\int_0^\infty dq \,q\,  \left[
(1+\ln (2aq))h_q'(0)-\int_0^\infty dz\, \ln z\, h_q^{''}(z)  \right]\,.
\label{21E_1}
\ee
In both expressions, the integrations are well convergent and deliver certain smooth functions of the plasma frequency $\om_p$. We show $\tilde{E}_1$ as a function of $\om_p$  in Fig. \ref{figDrudeTE} (left panel). For large $\om_p$
it decreases,
\be  \tilde{E}_1=\frac{\zeta(3)}{8\pi^2 a^3 \om_p}+O\left(\frac{1}{\om_p^2}\right)\,.
\label{21Etinf}\ee
It should be mentioned that this logarithmic contribution was already found in \cite{beze04-69-022119}, eq.(17).

The contribution from the TM mode is easier since in that case it is possible to expand  the integrand directly into powers of $\ga$ not producing singularities in the integrations (at least in first order in $\ga$). As a consequence, there is no logarithmic contribution and $E^{\rm TM}_1$ is a smooth function of $\om_p$ like $E_1^{\rm TE}$ and it can be calculated easily numerically. In this way, in \Ref{21E0} we have to insert $\tilde{E}_1$, \Ref{21Et}, and $E_1=E_1^{\rm TE}+E_1^{\rm TM}$.
%
%%%%%%%%%%%%%%%%%%%%%%%%%%%%%%%%%%%%%%%%%%%%%%%%%%%%%%%%%%%%%%%%%%%%%%%%%%%%%%%%%%%%%%%%%%%%
\subsection{The vacuum energy  for a dielectric with \DC conductivity}
The calculation of the vacuum energy for a dielectric with \DC conductivity goes in principle similar to that in the preceding subsection. In place of \Ref{21S0} we have now
\be S_0=-\frac{\pa E_0(\sigma)}{\pa \sigma}\,\frac{\pa \sigma}{\pa T}\,,
\label{22S0}
\ee
which  depends  on the  temperature  via the conductivity
$\sigma(T)$ in the permittivity \Ref{1dc} (we dropped the factor
$4\pi$ and defined $\sigma\equiv 4\pi \sigma_0$)
and we assume
\be \sigma(T)=\sigma_1 T^\al+\dots
\label{23Tsig}\ee
with $\al>0$.
Again we will
observe a logarithmic contribution for $T\to0$. This time it comes
from the TM mode and we note a formula in parallel to \Ref{21E0},
\be E_0(\sigma)=E_0(0)+\sigma\left(-\ln(2a\sigma)\,\tilde{E}_1+E_1\right)+\dots\,,
\label{22E0}
\ee
for small $\sigma$.
Starting from here, for technical reasons, we proceed in a slightly different way.
We start from the TM contribution and consider its derivative,
\be     \frac{\pa}{\pa \sigma}E_0^{\rm TM}(\sigma)=
        \frac{1}{4\pi^2}\int_0^\infty dq\,q\,\int_0^q
        \frac{d\xi}{\xi}\,g\left(\frac{\sigma}{\xi},\xi,q\right)\,,
\label{22E1}
\ee
with the notation
\be g(z,\xi,q)=  \frac{\pa}{\pa z}
\ln\left[1-\left( \frac{\left(\ep_0+z\right)q-\sqrt{(\ep_0-1+z)\xi^2+q^2}}
     {\left(\ep_0+z\right)q+\sqrt{(\ep_0-1+z)\xi^2+q^2}}
            \right)^2\,e^{-2aq}  \right]\,.
\label{22f}
\ee
In \Ref{22E1} we integrate by parts in the variable $\xi$,
\be     \frac{\pa}{\pa \sigma}E_0^{\rm TM}(\sigma)=
        \frac{1}{4\pi^2}\int_0^\infty dq\,q\,
        \left\{\ln (2aq)\ g\left(\frac{\sigma}{q},q,q\right)-
        \int_0^q d\xi\,\ln (2a\xi)\ \frac{\pa}{\pa \xi}
        \ g\left(\frac{\sigma}{\xi},\xi,q\right)    \right\}\,,
\label{22E2}
\ee
where we used that $g\left({\sigma}/{\xi},\xi,q\right)$ vanishes for $
\xi\to0$.

The integral in the figure brackets in this equation requires a special treatment. We divide it into two parts, $A_1$ and $A_2$, according to the division of the integration region into $\xi\in[0,\sqrt{\sigma/a}\,]$ and $\xi\in[\sqrt{\sigma/a},q]$. This is possible for any fixed $q$ since we are interested in the limit $\sigma\to0$. So we have
\be A_1=\int_0^{\sqrt{\sigma/a}}d\xi\, \ln(2a\xi) \,\frac{\pa}{\pa \xi}
        \ g\left(\frac{\sigma}{\xi},\xi,q\right) \,
\label{22A1}\ee
and $A_2$ accordingly.
In this integral we change the integration variable for  $\zeta=\xi/\sigma$,
\be A_1=\int_0^{\sqrt{1/a\sigma}}d\zeta\, \ln(2a\sigma\zeta) \,\frac{\pa}{\pa \zeta}
        \ g\left(\frac{1}{\zeta},\sigma\zeta,q\right) \,.
\label{22A2}\ee
Now it is possible to tend $\sigma\to0$ and we get
\be A_1=\int_0^{\infty}d\zeta\, \ln(2a\sigma\zeta) \,\frac{\pa}{\pa \zeta}
        \ g\left(\frac{1}{\zeta},0,q\right)+\dots \,,
\label{22A3}\ee
where the dots denote higher powers in $\sigma$.
In the contribution proportional to the logarithm of $\sigma$, the integration can be carried out using the derivative,
\be A_1=\ln(2a\sigma)\,g(0,0,q)+
\int_0^{\infty}d\zeta\, \ln \zeta  \ \frac{\pa}{\pa \zeta}
        \ g\left(\frac{1}{\zeta},0,q\right)+\dots \,,
\label{22A4}\ee
with
\be g(0,0,q)=\frac{4}{\ep_0^2-1}\left(e^{2aq}-r_0^{-2}\right)^{-1}
\label{22g}\ee
and $r_0$ is the reflection coefficient \Ref{2r0}. Since this is the only place where a logarithm of $\sigma$ appears we can now write down $\tilde{E}_1$ in \Ref{22E0}. With \Ref{22E2} and the above two formulas the integration over $q$ can be done and we obtain
\be \tilde{E}_1= \frac{1}{4\pi^2} \frac{{\rm Li}_2(r_0^2)}{(1-\ep_0^2)a^2}\,,
\label{22Et}
\ee
where ${\rm Li}_2$ is the polylogarithm. Next we need to calculate $A_2$. This is quite easy since we can put $\sigma=0$ directly in the integrand,
\be A_2=\int_0^{q}d\xi\, \ln(2a\xi) \,\frac{\pa}{\pa \xi}
        \ g\left(0,\xi,q\right) \,.
\label{22A5}\ee
Collecting all contributions we get from \Ref{22E2} and \Ref{22E1} the contribution from the TM mode to $E_1$,
\bea E_1^{\rm TM}&=&\frac{1}{4\pi^2}\int_0^\infty dq\,q\,
    \left\{\ln (2aq)\,g(0,q,q)
    -\int_0^\infty d \zeta\,\ln \zeta\,\frac{\pa}{\pa\zeta}\ g\left(\frac{1}{\zeta},0,q\right)
    \right.\nn\\&&\left.~~~~~~~~~~~~~~~~~~~~~~~~~~
    -\int_0^\infty d \xi\,\ln(2a \xi)\,\frac{\pa}{\pa\xi}\ g\left(0,\xi,q\right)  \right\}\,.
\label{22E1TM}
\eea
The integrals in this expression are convergent and deliver a smooth function of $\ep_0$ which can be evaluated numerically.
The contribution from the TE mode does not have a logarithmic term and it can be calculated by simply putting $\sigma=0$ in a formula in parallel to \Ref{22E1}. The emerging integrals are finite too. We restrict ourselves here with a representation of $\tilde{E}_1$ as a function of $\ep_0$ for $a=1$ in Fig. \ref{figDCTM} (right panel).

In this way, we observe in both models a logarithmic contribution to the vacuum energy for small parameter, $\ga$ or $\sigma$. If we insert these, together with \Ref{21Tal} resp. \Ref{23Tsig}, into \Ref{21S0} resp. \Ref{22S0}, we obtain, say for the Drude model,
\be \S_0=-\al\ga_1T^{\al-1}\left((\al \ln T+1)\tilde{E}_1+E_1\right)+\dots\,,
\label{23S0}\ee
and a similar formula for \DC conductivity. We observe not only a non zero, but even a diverging contribution to the entropy at vanishing temperature for $0<\al\le1$.
%
%%%%%%%%%%%%%%%%%%%%%%%%%%%%%%%%%%%%%%%%%%%%%%%%%%%%%%%%%%%%%%%%%%%%%%%%%%%%%%%%%%%%%%%%%%%%
\subsection{The temperature dependent part of the free energy  in the Drude model}
We start the calculation of the temperature dependent part of the
free energy from eqs. \Ref{2dF} and \Ref{2Fi} with the function $\varphi(\xi)$,
eq. \Ref{2fi}, with $\xi=ix$ inserted. So we have to consider
\be \varphi(ix)=\int_0^\infty dk\,k\,
\sum_{i=\TE,\TM}\ln\left(1-r_{i}^2\,e^{-2aq}\right),
\label{23fi0}
\ee
with $q=\sqrt{k^2-x^2}$. We divide the integration  into a first region, $k\in[0,x]$, and a second region, $k\in[x,\infty)$. We need the function $\varphi(ix)$ for small $x\sim T$. Therefore, in the first part of the integration region we have $k\le T$ and a factor $\sim T^2$ from $dk\,k$. As a consequence, the contribution from this region is by two additional powers of $T$ suppressed as compared with the second region where the integration region is infinite. Hence a contribution to the linear in T term can come from the second region only. In that region we change the variable of integration from $k$ for $q=\sqrt{k^2-x^2}$ (which is real)
and arrive at a representation of the temperature dependent part
of the free energy, up to higher orders in $T$, given by eqs.
\Ref{2dF}, \Ref{2BF} and \Ref{2Fi} with a function
\be  \varphi(ix)=\int_0^\infty dq\,q\,\sum_{i={\rm TE, TM}}\ln\left(1-r_i^2\, e^{-2aq}\right).
\label{23fi}
\ee
The reflection coefficients  are still given by eqs. \Ref{2rTE} and \Ref{2rTM} but with the substitution $\xi_l\to ix$ and with the permittivities \Ref{1D} and \Ref{1dc} with $\xi\to ix$.

Now we consider the Drude model. For small $T$, the leading contribution results from the TE mode. With \Ref{1D} and \Ref{2rTE} we get
\be \Delta_T\F^{\rm Drude, TE}=\frac{1}{4\pi^2}\int_0^\infty \frac{d x}{e^{x/T}-1}
    \int_0^\infty dq\,q\,
    i\left(h_q\left(\frac{\ga}{i x}\right)-h_q\left(\frac{\ga}{-i x}\right)\right)+\dots
\label{23FD}
\ee
with the function $h_q(y)$ defined in \Ref{21fq}. Now we consider   $\al\ge1$ in \Ref{21Tal}, i.e., $\ga$ decreases not slower than the first power of the temperature. We make the substitution $x=\ga\zeta$,
\be \Delta_T\F^{\rm Drude, TE}=\frac{1}{4\pi^2}\int_0^\infty d \zeta \frac{\ga}{e^{\ga \zeta/T}-1}
    \int_0^\infty dq\,q\,
    i\left(h_q\left(\frac{1}{i \zeta}\right)-h_q\left(\frac{1}{-i \zeta}\right)\right)
     \nn \\
  %  &&\times i\left[
%    \ln\left(1-\left(\frac{q-\sqrt{\om_p^2\frac{i x}{1+ix}+q^2}}
%                    {q+\sqrt{\om_p^2\frac{i x}{1+ix}+q^2}}\right)^2\,e^{-2aq}\right)
%                    -{\rm c.c.}\right]\,,
\label{23FD1}
\ee
and note the ratio $\ga/T$ in the Boltzmann factor. Now we can tend $T\to0$ in the integrand and note \Ref{21Tal}. For $\al=1$ we get
\be \Delta_T\F^{\rm Drude, TE}=\frac{T}{16\pi a^2}\, f^{\rm D}(\ga_1)
\label{23dF}
\ee
with
\be f^{\rm D}(\ga_1)=  (2a)^2
\int_0^\infty d \zeta \frac{\ga_1}{e^{\ga_1 \zeta}-1}
    \int_0^\infty dq\,q\,
    i\left(h_q\left(\frac{1}{i \zeta}\right)-h_q\left(\frac{1}{-i \zeta}\right)\right)\,.
    %\nn \\
%    &&\times i\left[
%    \ln\left(1-\left(\frac{q-\sqrt{2a\om_p^2\frac{i x}{1+ix}+q^2}}
%                    {q+\sqrt{2a \om_p^2\frac{i x}{1+ix}+q^2}}\right)^2\,e^{-q}\right)
%                    -{\rm c.c.}\right]\, ,
\label{23fD}
\ee
 This is a smooth function of $\ga_1$ and of $2a\om_p$. For $\al>1$, because of
 \be \frac{\ga}{e^{\ga\zeta/T}-1}=\frac{T}{\zeta}+\dots
 \label{23to0}\ee
for $T\to0$, we get from \Ref{23FD1} the linear in $T$ term \Ref{23dF} with the same function $f^{\rm D}$, however with zero argument, $f^{\rm D}(0)$. Its explicit expression reads
\bea f^{\rm D}(0)&=&
\int_0^\infty  \frac{d \zeta}{\zeta}
    \int_0^\infty d\tilde{q}\,\tilde{q}\,
    \nn \\
    &&\times i\left[
    \ln\left(1-\left(\frac{\tilde{q}-\sqrt{(2a\om_p)^2\frac{i \zeta}{1+i\zeta}+\tilde{q}^2}}
                    {\tilde{q}+\sqrt{(2a \om_p)^2\frac{i \zeta}{1+i\zeta}+\tilde{q}^2}}\right)^2\,e^{-\tilde{q}}\right)
                    -{\rm c.c.}\right]\,,
\label{23fD0}
\eea
where we inserted the explicit expression \Ref{21fq} and substituted $q=\tilde{q}/2a$.
Here, and below, we use the notation 'c.c.'  indicating that the complex conjugate inside the square bracket must be subtracted.  We would like to comment on the know property of the linear in $T$ term not to depend on the relaxation parameter. This property holds for $\al>1$ and it results from the ratio in Eq.\Ref{23to0} having a smooth limit for $T\to0$ which does not depend on $\gamma_1$.

In the other case of a decreasing relaxation coefficient where $0<\al<1$ holds, we start from eq.\Ref{23FD} and make the substitutions $x=T\zeta$ and $q=\tilde{q}\sqrt{T/\ga}$,
\be \Delta_T\F^{\rm Drude, TE}=\frac{T^2}{4\pi^2\ga}\int_0^\infty d \zeta \frac{\ga}{e^{ \zeta}-1}
    \int_0^\infty d\tilde{q}\,\tilde{q}\,
    i\left(h_{\tilde{q}\sqrt{T/\ga}}\left(\frac{\ga}{i \zeta T}\right)-h_{\tilde{q}\sqrt{T/\ga}}\left(\frac{\ga}{-i \zeta T}\right)\right)\,.
\label{23FD2}
\ee
In this representation it is possible to tend $T\to0$ in the integrand. The emerging integrals are finite. However, the factor in front, $T^2/\ga\sim T^{2-\al}$ goes to zero faster than a first power of $T$. Hence it does not contribute to the violation of the 3rd law.

\begin{figure}[h]
\hspace{0.51cm}\includegraphics[width=7cm]{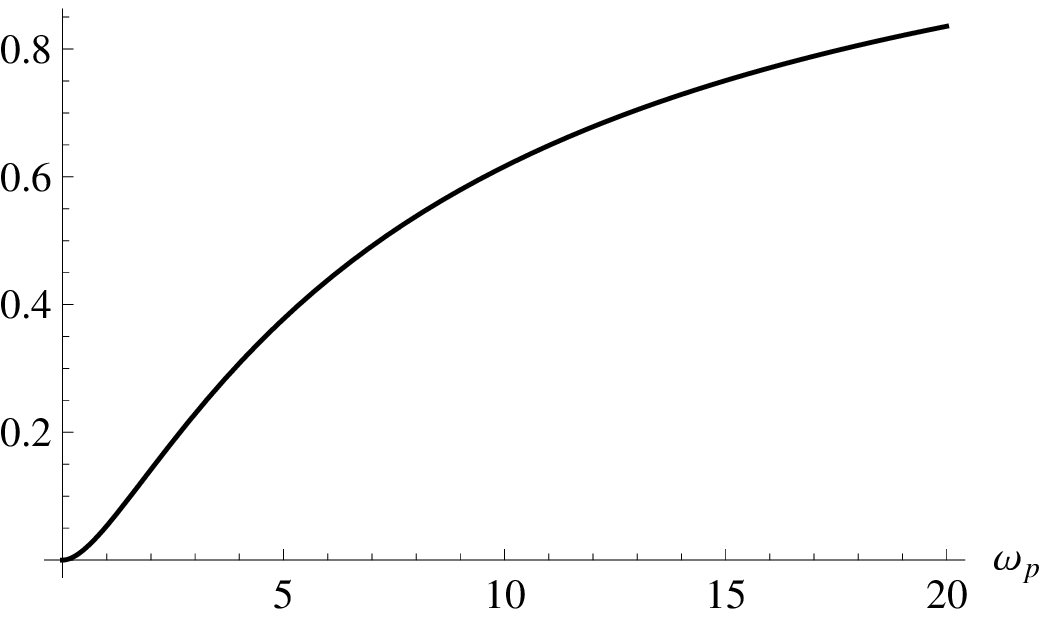}
\hspace{0.21cm}\includegraphics[width=7cm]{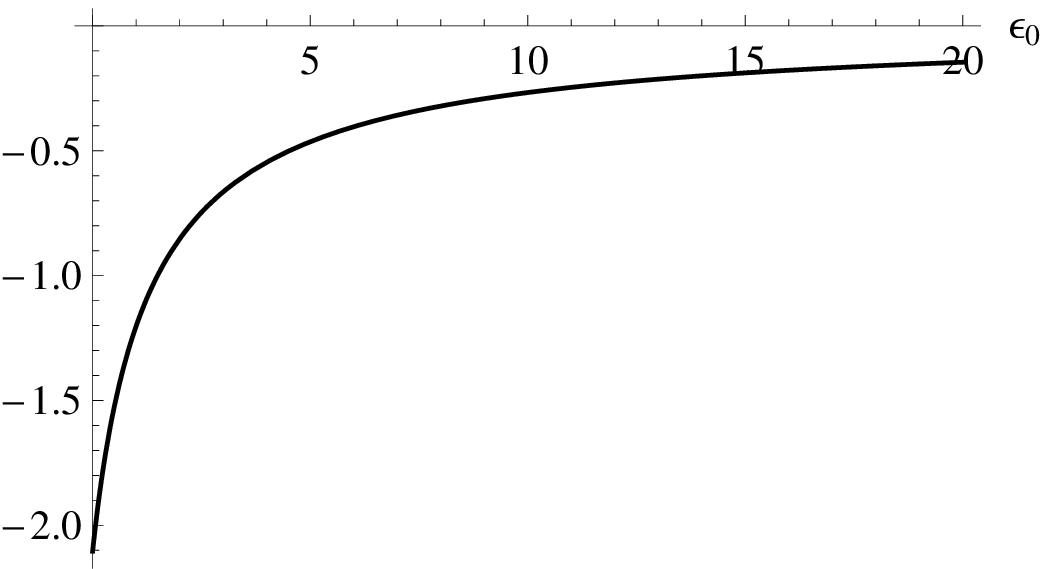}
\caption{The dependence on $\om_p$ for $a=1$ of the function $f^{\rm D}(0)$, Eq. \Ref{23fD0}, (left panel) and the dependence on $\ep_0$ of the function $f^{\DC}(0)$, Eq. \Ref{24fdc0}, (right panel).}
\label{fig||fD}
\end{figure}

It is easy to see that $f^{\rm D}(0)$, Eq.\Ref{23fD0}, is a smooth function of $2a\om_p$. It is also
possible to show that this is just the same
function as $f^{\rm D}_{\rm lin}$, Eq. \Ref{2DLin} in \Ref{2FD}: $f^{\rm D}(0)=f^{\rm D}_{\rm lin}$. We have plotted $f^{\rm D}(0)$ as a function of $\om_p$ (with $a=1$) in Fig. \ref{fig||fD} (left panel).

In this way, for a decrease of the
relaxation parameter $\ga(T)$ linear in $T$ or faster we have a
linear contribution to the free energy. If the decrease of
$\ga(T)$ is slower this linear term disappears. Finally we remind that the TM mode
does not contribute to a linear term.

For a better understanding of the structures involved and as
reference for the case of a sphere in front of the plane we remind
here shortly the case of a fixed relaxation parameter $\ga$. In
that case there is no linear term and $T^2$ is the leading order
which receives contributions from both, TE and TM modes. The
starting point is again eq. \Ref{23FD} with the function $h^{\rm TE}_q(z)\equiv h_q(z)$,
\Ref{21fq}, for the TE mode. The contribution from the TM mode is
given by the same formula, \Ref{23FD}, with
\be h^{\rm TM}_q(z)=\ln\left[ 1-
\left(\frac{\left(1+\frac{\om_p^2}{\xi^2(1+z)}\right)q-\sqrt{\frac{\om_p^2}{1+z}+q^2}}
{\left(1+\frac{\om_p^2}{\xi^2(1+z)}\right)q+\sqrt{\frac{\om_p^2}{1+z}+q^2}}\right)^2
\, e^{-2aq}\right] \,
\label{23HTM} \ee
instead. For the TE mode one needs to make the substitution $x=T\zeta$ and $q=\sqrt{T}\tilde{q}$. After that one can put $T=0$ directly in
the integrands and one comes to the expression
\be     \Delta_T\F^{\rm Drude, TE}=
        \frac{T^2}{4\pi^2}\int_0^\infty\frac{d\zeta}{e^\zeta-1}\, \int_0^\infty d\tilde{q}\,\tilde{q}\
        i\left\{\ln\left[1-
        \left(\frac{\tilde{q}-\sqrt{\om_p^2\frac{i\zeta}{\ga}+\tilde{q}^2}}
        {\tilde{q}+\sqrt{\om_p^2\frac{i\zeta}{\ga}+\tilde{q}^2}}
        \right)^2\right]
        -
        %\ln\left(1-
%        \left(\frac{q-\sqrt{\om_p^2\frac{-ix}{\ga}+q^2}}{q+\sqrt{\om_p^2\frac{ix}{\ga}+q^2}}
%        \right)^2\right)
{\rm c.c.}\right\}\,.
\label{23gaf1}\ee
These integrals can be computed easily delivering
\be  \Delta_T\F^{\rm Drude, TE}=
        \frac{2\ln2-1}{48}  \, \frac{\om_p^2\,T^2}{\ga }+\dots\,.
\label{23gaf2}\ee
We note that this expression is just the same as that which follows from \Ref{23fD} for large $\gamma_1$ with the formal substitution $\ga_1\to\ga/T$.

For the TM mode one starts from the same formula, \Ref{23FD}, now
with \Ref{23HTM} inserted. Here one needs only to substitute $x=T\zeta$ and one can expand the integrand for small $T$. This expansion
starts from a first order term which delivers
\be \Delta_T\F^{\rm Drude, TM}=
        \frac{T}{4\pi^2}\int_0^\infty\frac{d\zeta}{e^\zeta-1}\, \int_0^\infty dq\,q\
        \frac{-8\ga x T}{(e^{2aq}-1)\,\om_p^2}+\dots\,.
\label{23gaf3}\ee
Carrying out the integration one comes to
\be \Delta_T\F^{\rm Drude, TM}=
          -\frac{\pi^2}{18}  \, \frac{\ga\,T^2}{ (2a)^2 \om_p^2}+\dots\,,
\label{23gaf4}\ee
which must be added to \Ref{23gaf2}.
%
%%%%%%%%%%%%%%%%%%%%%%%%%%%%%%%%%%%%%%%%%%%%%%%%%%%%%%%%%%%%%%%%%%%%%%%%%%%%%%%%%%%%%%%%%%%%
\subsection{The temperature dependent part of the free energy   for a dielectric with \DC conductivity}
For a dielectric with \DC conductivity we apply the same scheme of
calculations as in  the preceding subsection. In this case the
interesting contribution comes from the TM mode and in place of
\Ref{23FD}, using \Ref{2rTM} and \Ref{1dc}, we have now
\be \Delta_T\F^{\rm \DC, TM}=\frac{1}{4\pi^2}\int_0^\infty \frac{d
x}{e^{x/T}-1}
    \int_0^\infty dq\,q\, i\left[
    \ln\left(1-r_{\rm TM}^2\,e^{-2aq}\right)
                    -{\rm c.c.}\right]\,
\label{24FD}
\ee
with
\be  r_{\rm TM}=
 \frac{\left(\ep_0(ix)+\frac{ \sigma}{ix}\right)q-\sqrt{\left(-\ep_0(ix)+1\right)x^2+ i x \sigma+q^2}}
           {\left(\ep_0(ix)+\frac{ \sigma}{ix}\right)q+\sqrt{\left(-\ep_0(ix)+1\right)x^2+ i x \sigma+q^2}}\,.
\label{24r}
\ee
We make the substitution $x=\sigma \zeta$ and consider $\sigma(T)$ decreasing according to \Ref{23Tsig} with $\al\ge 1$.
After this substitution we can tend $T\to0$ in the integrand. For $\al=1$ we come to
\be   \Delta_T\F^{\rm \DC, TM}=-\frac{T}{16\pi a^2}\,f^{\rm
\DC}(\sigma_1) \label{24dF} \ee
with
\bea f^{\rm \DC}(\sigma_1)&=& -\frac{(2a)^2}{4\pi^2}\int_0^\infty
d \zeta\frac{\sigma_1}{e^{\sigma_1 \zeta}-1}
    \int_0^\infty dq\,q\,
     i\left[
    \ln\left(1-\left(\frac{1+i\zeta(\ep_0-1)}{1+i\zeta(\ep_0+1)}\right)^2\,e^{-2aq}\right)
    \right.\nn \\ &&\left.
   ~~~~~~~~~~~~~~~~~~~~~~~~~~~~~~~~ -
     \ln\left(1-\left(\frac{1-i\zeta(\ep_0-1)}{1-i\zeta(\ep_0+1)}\right)^2\,e^{-2aq}\right)
                   \right]\,.
\label{24fdc}
\eea
This is a smooth function of $\sigma_1$ taking a finite  limiting
value for $\sigma_1=0$ and it decreases for $\sigma_1\to\infty$.
For $\al>1$ we come to the same expression \Ref{24dF}, but with the function  $f^{\DC}$ with zero argument,
\bea f^{\rm \DC}(0)&=&-\frac{1}{4\pi^2}\int_0^\infty \frac{d \zeta}{\zeta}
    \int_0^\infty d\tilde{q}\,\tilde{q}\,
     i\left[
    \ln\left(1-\left(\frac{1+i\zeta(\ep_0-1)}{1+i\zeta(\ep_0+1)}\right)^2\,e^{-\tilde{q}}\right)
      \right.\nn \\ &&\left.
   ~~~~~~~~~~~~~~~~~~~~~~~~~~ -
    \ln\left(1-\left(\frac{1-i\zeta(\ep_0-1)}{1-i\zeta(\ep_0+1)}\right)^2\,e^{-\tilde{q}}\right)
                    \right]\,.
\label{24fdc01}
\eea

So we see in this case a picture similar to the  Drude model. If
the conductivity $\sigma(T)$ decreases at least linear with $T$,
we have a linear in $T$ contribution to the temperature dependent
part of the free energy. If it decreases slower (that is, $\al<1$) this linear term can be shown to be absent.

In \Ref{24fdc01}, in the reflection coefficient, the $q$-dependence dropped out and
we substituted $q=\tilde{q}/2a$. This expression can be simplified by
an expansion of the logarithm into a sum after which the
$q$-integration can be done immediately,
\be
f^{\rm dc}(0)=-\frac{1}{4\pi^2}\sum_{k=1}^\infty\frac{-1}{k^3}\int_0^\infty \frac{d x}{x}\
        i\left[
    \left(\frac{1+ix(\ep_0-1)}{1+ix(\ep_0+1)}\right)^{2k}
    - \left(\frac{1-ix(\ep_0-1)}{1-ix(\ep_0+1)}\right)^{2k}   \right]\,.
\label{24fdc02}
\ee
Now the $x$-integration can be done too and we arrive at
\be f^{\rm dc}(0)={\rm Li}_3(r_0^2)-\zeta(3).
\label{24fdc0}
\ee
This is just the same function \Ref{2fDlin} as in eq.\Ref{2Fdc}, i.e.,
$f^{\rm DC}(0)=f_{\rm lin}^{\rm DC}$
holds. We show it on Fig. \Ref{fig||fD} (right panel).
Since in this case the TE mode does not contribute we reproduced the linear term \Ref{2FD}.

We conclude this subsection by comparing with the case of a fixed
DC conductivity $\sigma$.  The temperature dependent part of the
free energy is given by eq. \Ref{23FD} with the functions
\be h^{\rm TE}_q(\sigma,x)=\ln\left[
        1-\left(
        \frac
        {q-\sqrt{\left(\ep_0-1+\frac{\sigma}{x}\right)x^2+q^2}}
        {q+\sqrt{\left(\ep_0-1+\frac{\sigma}{x}\right)x^2+q^2}}
        \right)^2e^{-2aq}
        \right]
\label{24hTE}
\ee
and
\be h^{\rm TM}_q(\sigma,x)=\ln\left[
        1-\left(
        \frac
        {\left(\ep_0+\frac{\sigma}{x}\right)q-\sqrt{\left(\ep_0-1+\frac{\sigma}{x}\right)x^2+q^2}}
        {\left(\ep_0+\frac{\sigma}{x}\right)q+\sqrt{\left(\ep_0-1+\frac{\sigma}{x}\right)x^2+q^2}}
        \right)^2e^{-2aq}
        \right]
\label{24hTM} \ee
for the two modes. In both cases one needs to make the
substitution $x= T\zeta$. In the TE case one needs to integrate by
parts for $q$ several times to see that the TE mode results in a
order $T^4$ contribution and can be dropped. In the TM
contribution one can expand
\be     h^{\rm TM}_q(\sigma,\zeta T)=
        \frac{8 \zeta T}{\sigma (1-e^{2aq})}+\dots
\label{24caf1} \ee
and carry out the remaining integrations. The result is simply

\be     \Delta_T\F^{TM}=-\frac{\pi^2}{18}\,\frac{T^2}{(2a)^2\sigma}+\dots\,,
\label{24caf2}
\ee
which is the counterpart  of \Ref{23gaf4}. Moreover, \Ref{24caf2} and \Ref{23gaf4} are related by the substitution $\sigma\to\ga/\om_p^2$ which just relates $\ep^{\rm D}$, \Ref{1D}, with $\ep^{\DC}$, \Ref{1dc}, in leading order for $\xi\to0$ (note our notation $\sigma \equiv 4\pi\sigma_0$).

It should be mentioned that the low temperature behavior for fixed parameters in known. It was calculated earlier using other methods. For example, Eq. \Ref{23gaf2} can be found in \cite{hoye07-75-051127} and Eq. \Ref{24caf2} in \cite{elli08-78-021117}.
%%%%%%%%%%%%%%%%%%%%%%%%%%%%%%%%%%%%%%%%%%%%%%%%%%%%%%%%%%%%%%%%%%
\section{The free energy for a sphere in front of a plane}
In this section we extend the results of the preceding section to the geometry of a sphere, metallic or dielectric,  in front of a conducting plane. The radius of the sphere is $R$ and and the distance from its center to the plane  is $L$.
Technically, this is an extension of our previous papers,  \cite{bord10-1007} and \cite{bord10-81-085023}. So for the temperature dependent part of the free energy  we use the same basic formula,
\be \Delta_T\F=\frac{1}{2\pi}\int_0^\infty dx\, n_T(x)\
    i\Tr\left[ \ln \left(1-\mathbb{M}(ix)\right)-\ln\left(1-\mathbb{M}(-ix)\right)\right]\,,
\label{3F}
\ee
with the Boltzmann factor given by Eq.\Ref{2BF}. The matrix $\mathbb{M}(\xi)$ has the entries
\be  {M}_{l,l'}
=
\sqrt{\frac{\pi}{4\xi L}}\sum_{l''=|l-l'|}^{l+l'}K_{\nu''}(2\xi
L) H_{ll'}^{l''}\,
 \left(\begin{array}{cc}\Lambda_{l,l'}^{l''}&\tilde{\Lambda}_{l,l'}
\\ \tilde{\Lambda}_{l,l'}&\Lambda_{l,l'}^{l''}\end{array}\right)
\left(\begin{array}{cc}d^{\rm TE}_l(\xi R)&0
\\ 0&-d^{\rm TM}_l(\xi R)\end{array}\right)
 \label{3M}
\ee
with the coefficients $ H_{ll'}^{l''}$ and $\Lambda_{l,l'}^{l''}$ given by eqs.(7) and (10) in \cite{bord10-1007} and it is a matrix with respect to the polarizations. For the TE mode the function $d_l^{\rm TE}(\xi)$ is given by
\be d^{\rm TE}_l(\xi)=\frac{2}{\pi}\,
        \frac   {\sqrt{\ep}s_l(\xi)s'_l(n\xi)-\sqrt{\mu}s'_l(\xi)s_l(n\xi)}
                {\sqrt{\ep}e_l(\xi)s'_l(n\xi)-\sqrt{\mu}e'_l(\xi)s_l(n\xi)}\,,
\label{3tTE}
\ee
where $n=\sqrt{\ep \mu}$ is the refraction index. The function $d_l^{\rm TM}(\xi)$  for the TM mode follows by interchanging $\ep$ and $\mu$.

These functions are expressed in terms of the know modified spherical Bessel functions,
$s_l(z)=\sqrt{\pi z/2}I_{l+1/2}(z)$ and $e_l(z)=\sqrt{2 z/\pi}K_{l+1/2}(z)$. For the following it is useful to separate the powers of the argument in front of the ascending series. We represent
\be \begin{array}{rclrrcl}
    s_l(z)&=&\sqrt{\pi}\left(\frac{z}{2}\right)^{l+1}i_l(z), &~~~~~~~~~
     e_l(z)&=&\sqrt{\pi}\left(\frac{z}{2}\right)^{l+1}k_l(z), \\[7pt]
      z\frac{\pa}{\pa z}\,s_l(z)&=&\sqrt{\pi}\left(\frac{z}{2}\right)^{l+1}\tilde{i}_l(z), &
      z\frac{\pa}{\pa z}\,e_l(z)&=&\sqrt{\pi}\left(\frac{z}{2}\right)^{l+1}\tilde{k}_l(z),
\end{array}
\label{3ik1}
\ee
with
\be \begin{array}{rclrrcl}
    i_l(z)&=&\left(\frac{z}{2}\right)^{-l-1/2}I_{l+1/2}(z), &~~~~~~~~~
     k_l(z)&=&\frac{2}{\pi}\left(\frac{z}{2}\right)^{l+1/2}K_{l+1/2}(z) ,\\[7pt]
      \tilde{i}_l(z)&=&\left(l+1+z\frac{\pa}{\pa z}\right){i}_l(z), &
       \tilde{k}_l(z)&=&\left(-l+z\frac{\pa}{\pa z}\right){k}_l(z).
\end{array}
\label{3ik2}
\ee
All these functions  have power series expansions, e.g.,
\be
i_l(z)=i_l(0)+i_l^{(1)}z^2+\dots\,,~~~~~\tilde{i}_l(z)=\tilde{i}_l(0)+\tilde{i}_l^{(1)}z^2+\dots\,,
\label{3i1k1}\ee
which we will need below.

These definitions allow to rewrite the functions $d_l^{\rm TX}(\xi)$ in the form
\be d^{\rm TX}_l(\xi)=\frac{2}{\pi}\left(\frac{\xi}{2}\right)^{2l+1}t^{\rm TX}_l(\xi)
\label{3itTX}
\ee
with
\be t^{\rm TE}_l(\xi)=\frac{i_l(\xi)\tilde{i}_l(\sqrt{\ep} \xi)-\tilde{i}_l(\xi)i_l(\sqrt{\ep}\xi)}
                    {k_l(\xi)\tilde{i}_l(\sqrt{\ep} \xi)-\tilde{k}_l(\xi)i_l(\sqrt{\ep}\xi)}
\label{3tTE1}
\ee
and
\be t^{\rm TM}_l(\xi)=
\frac{\frac{1}{\sqrt{\ep}}i_l(\xi)\tilde{i}_l(\sqrt{\ep} \xi)-\tilde{i}_l(\xi)i_l(\sqrt{\ep}\xi)}
    {\frac{1}{\sqrt{\ep}}k_l(\xi)\tilde{i}_l(\sqrt{\ep} \xi)-\tilde{k}_l(\xi)i_l(\sqrt{\ep}\xi)}\,,
\label{3tTM1}
\ee
where we  restricted ourselves the a pure dielectric ball putting $\mu=1$. Now the entries of the matrix $\mathbb{M}$ take the form
\be  {M}_{l,l'}
=
\frac{\sqrt{\pi}}{2}\sum_{l''=|l-l'|}^{l+l'}
    \frac{(\xi R/2)^{l+l'+1}}{(\xi L)^{l''+1}} \, k_{l''}(2\xi L) \,
     H_{ll'}^{l''}\,
 \left(\begin{array}{cc}\Lambda_{l,l'}^{l''}&\tilde{\Lambda}_{l,l'}
    \\ \tilde{\Lambda}_{l,l'}&\Lambda_{l,l'}^{l''}\end{array}\right)
\left(\begin{array}{cc}t^{\rm TE}_l(\xi R)&0
\\ 0&-t^{\rm TM}_l(\xi R)\end{array}\right)\,.
 \label{3M1}
\ee
We redistributed a factor $(\xi R/2)^{l'}$ which is admissible under the trace in \Ref{3F}.

For the low temperature expansion this expression can be simplified. Below we need the lowest orders for $\xi\to0$,
\be \mathbb{M}=\mathbb{M}_0+\mathbb{M}_1\xi+\dots\,.
\label{3M2}
\ee
From the powers of $\xi$ in \Ref{3M1} it follows that only $l''=l+l'$ contributes to \Ref{3M1} and that $\mathbb{M}_0$ is diagonal in the polarizations. The latter follows form the additional factor of $\xi$ in $\tilde{\Lambda}_{l,l'}^{l''}$. For this reason, and because of the trace in \Ref{3F}, the temperature dependent part of the free energy becomes a sum of the two polarizations,
\be \Delta\F=\Delta\F^{\rm TE}+\Delta\F^{\rm TM}+\dots\,,
\label{3F1}
\ee
which holds in the orders of $T$ we are interested in.
We continue with a separate consideration of the Drude model and the \DC conductivity in the following two subsections.
%%%%%%%%%%%%%%%%%%%%%%%%%%%%%%%%%%%%%%%%%%%%%%%%%%%%%%%%%%%%%%%%%%%%%
\subsection{Ball described by the Drude model}
In this subsection we consider a ball described by the Drude model in front of a conducting plane. It turns out that we can act in quite close analogy to the planar case in subsection 2.3. We assume  a relaxation parameter decreasing with temperature  according to \Ref{21Tal}. We make the substitutions $x=\ga\zeta$ in \Ref{3F} and with \Ref{1D} we note
\be    \frac{1}{\ep^{\rm D}(\ga\zeta)}=\frac{\ga\sqrt{\zeta(1+\zeta)}}{\om_p}+\dots~~~~~
\mbox{and}~~~~~\sqrt{\ep^{\rm D}(\ga\zeta)}\ \ga\zeta=\om_p\sqrt{\frac{\zeta}{1+\zeta}}+\dots\,
\label{31ep}
\ee
for $\ga\to0$. After that we expand the matrix $\mathbb{M}$ for small $\ga$. First we expand $t^{\rm TE}(\xi)$.
With \Ref{31ep} we get from \Ref{3tTE1} for the TE mode
\be {t^{\rm TE}(\ga\zeta)}= t^{\rm TE}_0(\zeta,\om_p) +\dots
\label{31a}\ee
with
\be   t^{\rm TE}_0(\zeta,\om_p) =
\frac
{i_l(0)\, \tilde{i}_l\left(\om_p\sqrt{\frac{\zeta}{1+\zeta}}\,\right)-\tilde{i}_l(0)\, {i}_l\left(\om_p\sqrt{\frac{\zeta}{1+\zeta}}\,\right)}
{k_l(0)\, \tilde{i}_l\left(\om_p\sqrt{\frac{\zeta}{1+\zeta}}\,\right)-\tilde{k}_l(0)\, {i}_l\left(\om_p\sqrt{\frac{\zeta}{1+\zeta}}\,\right)} \,,
\label{31tTE2}
\ee
whereas for the TM mode
\be t^{\rm TM}(\ga\zeta)=O(\ga^2)
\label{31TM2}
\ee
holds. So, only the TE mode contributes to the linear in $T$ term. In the remaining factors in $\mathbb{M}$ we can put also $\ga=0$ and come to
\bea M_{l,l'}^{\rm Drude}(\zeta)&\equiv& M_{l,l'}(\ga\zeta)_{|_{\ga=0}}\,,\nn\\
    &=&  \frac{\sqrt{\pi}}{2}\left(\frac{\eps}{2}\right)^{2l+1} k_{l+l'}(0)H^{l+l'}_{l,l'}
\Lambda^{l+l'}_{l,l'}t^{\rm TE}_0(\zeta,\om_p R)\,,
\label{31M3}\eea
where we defined
\be     \eps=\frac{R}{L}
\label{31eps}
\ee
for the ratio of radius of the ball to the distance between the plane and the center of the ball. It must be mentioned that $M_{l,l'}^{\rm Drude}(\zeta)$ still depends on $\zeta$ through $t^{\rm TE}_0$ because of \Ref{31ep}.

\begin{figure}[t]
\hspace{1cm}\includegraphics[width=7cm]{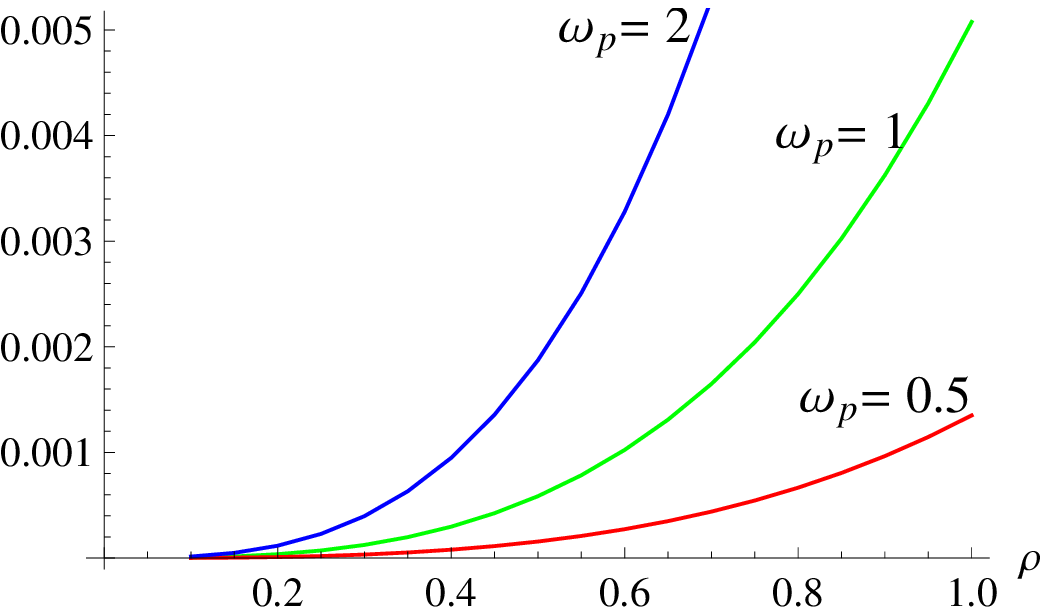}\includegraphics[width=7cm]{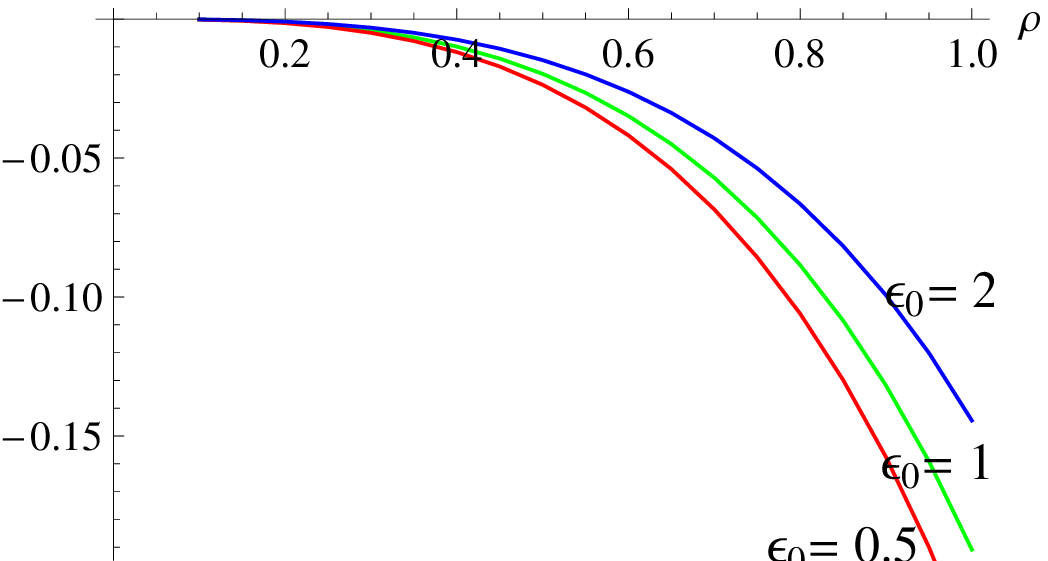}
\caption{The function $ f^{\rm D}_{\rm ball}(\eps,\om_p)$, \Ref{31f},    as a function of $\rho$ for several values of $\om_p$ (left panel) and the function $ f^{\DC}_{\rm ball}(\eps,\ep_0)$, \Ref{32f},    as a function of $\rho$ for several values of $\ep_0$ (right panel).}
\label{fig3Drude}\label{fig3DC}
\end{figure}

Now we insert $M_{l,l'}^{\rm Drude}(\zeta)$, \Ref{31M3}, into \Ref{3F},
\be \Delta_T\F=\frac{1}{2\pi}\int_0^\infty d\zeta\, \frac{\ga}{e^{\ga \zeta/T}-1}\
    i\Tr\left[  \left(1-\mathbb{M}^{\rm Drude}(i\zeta)\right)-\left(1-\mathbb{M}^{\rm Drude}(-i\zeta)\right)\right]\,.
\label{31F}
\ee
This formula is in parallel to \Ref{23dF} in the planar case. It has a non vanishing limit for $T\to0$ in case of a $\ga$ decreasing not slower than linear, i.e., for $\al\ge1$ in \Ref{21Tal}. Now we restrict ourselves to  $\al>1$  and come to
\be\Delta_T\F=\frac{T}{2\pi}\,f^{\rm D}_{\rm ball}(\eps,\om_pR)+O(T^2)
\label{31F1}
\ee
with
\be f^{\rm D}_{\rm ball}(\eps,\om_pR)=
    \frac{1}{2\pi}\int_0^\infty   \frac{d\zeta}{\zeta}\
    i\Tr\left[  \left(1-\mathbb{M}^{\rm Drude}(i\zeta)\right)-\left(1-\mathbb{M}^{\rm Drude}(-i\zeta)\right)\right]\,.
\label{31f}
\ee
This function describes the contribution linear in $T$ in the Drude model. In Fig.\ref{fig3Drude} we have plotted $f^{\rm D}_{\rm ball}(\eps,\om_p)$ as function of $\eps$ for several values of $\om_p$. For the calculation of the trace one needs to make a truncation of the orbital momenta, $l\le l_m$. In this case it turned out that for all values of the parameters $\eps$ and $\om_p$ a few lowest $l\lesssim 4$ are sufficient. This includes, for instance, the case $\eps=1$, i.e., zero separation.

%%%%%%%%%%%%%%%%%%%%%%%%%%%%%%%%%%%%%%%%%%%%%%%%%%%%%%%%%%%%%%%%%%%%%
\subsection{Dielectric ball with DC conductivity}
In this subsection we derive the linear contribution for a dielectric ball with DC conductivity. We use the permittivity \Ref{1dc} and  assume   a decrease of the conductivity according to \Ref{23Tsig}.  We substitute $\xi= \sigma\zeta$ and note
\be \frac{1}{\ep^{\DC}(\sigma\zeta)}={\frac{\zeta}{\ep_0+\zeta}}+\dots, ~~\mbox{and}~~~
    \sqrt{\ep^{\DC}(\sigma\zeta)}\,\sigma\zeta R=\sigma\sqrt{\zeta(\zeta+\ep_0)}R+\dots\,.
\label{32}
\ee
We have to insert these expansions into $t^\TE_l(\xi)$, \Ref{3tTE1}, and $t^\TM_l(\xi)$, \Ref{3tTM1}. For the TE case we observe that $t^\TE_l(\sigma\zeta)$ is of higher order in $\sigma$ and, for this reason, it does not contribute to the linear in $T$ term. In opposite, for the TM mode we have
\bea t^\TM_0(\zeta,\sigma)&\equiv&{t^{\TM}_l(\sigma\zeta)}_{|\sigma=0}
    \nn\\&=&    \frac{\left(\frac{\zeta}{\ep_0+\zeta}-1\right)i_l(0)\tilde{i}_l(0)}
                    { \frac{\zeta}{\ep_0+\zeta} k_l(0)\tilde{i}_l(0)-\tilde{k}_l(0)i_l(0)}\,,
\label{321}
\eea
which is the analog of eq.\Ref{31tTE2} for the Drude model.
The remaining calculations go in parallel to the preceding subsection. In the remaing parts of the matrix $\mathbb{M}$ we put $\sigma=0$ and come to
\bea M_{l,l'}^{\DC}(\xi)=
  \frac{\sqrt{\pi}}{2}\left(\frac{\eps}{2}\right)^{2l+1} k_{l+l'}(0)H^{l+l'}_{l,l'}
\Lambda^{l+l'}_{l,l'}t^{\rm TM}_0(\zeta,\sigma)\,.
\label{32M3}\eea
Now we insert this $\mathbb{M}_{l,l'}^{\DC}(\xi)$ into \Ref{3F} and get
\be \Delta_T\F=\frac{1}{2\pi}\int_0^\infty d\zeta\, \frac{\sigma}{e^{\sigma \zeta/T}-1}\
    i\Tr\left[  \left(1-\mathbb{M}^{\DC}(i\zeta)\right)-\left(1-\mathbb{M}^{\DC}(-i\zeta)\right)\right]\,.
\label{32F}
\ee
In this way, also for dc conductivity we have a linear contribution for $T\to0$ in case $\sigma$ vanishes linear in $T$. If it vanishes faster, just like before, the linear term becomes independent form $\sigma$ and it is
\be\Delta_T\F=\frac{T}{2\pi}\,f^{\DC}_{\rm ball}(\eps,\ep_0)+O(T^2)
\label{32F1}
\ee
with
\be f^{\DC}_{\rm ball}(\eps,\ep_0)=
    \int_0^\infty   \frac{d\zeta}{\zeta}\
    i\Tr\left[  \left(1-\mathbb{M}^{\DC}(i\zeta)\right)-\left(1-\mathbb{M}^{\DC}(-i\zeta)\right)\right]\,.
\label{32f}
\ee
This function can be calculated in the same way as $f^{\rm D}_{\rm ball}(\eps,\om_p)$ in the preceding subsection. We have plotted $f^{\DC}_{\rm ball}(\eps,\ep_0)$ in Fig. \ref{fig3DC} as function of $\eps$ for several values of $\ep_0$. We used a truncation $l\le l_m$ of the orbital momenta and again some low $l$ were sufficient.
%
%%%%%%%%%%%%%%%%%%%%%%%%%%%%%%%%%%%%%%%%%%%%%%%%%%%%%%%%%%%%%%%%%%%%%
\subsection{Both models with fixed parameters}
In this subsection we consider both models, Drude model and DC conductivity, with fixed parameters. The starting point is again Eq.\Ref{3F} for the temperature dependent part of the free energy. For the limit $T\to0$, we need the matrix $\mathbb{M}(\xi)$ for small $\xi$. First we consider the permittivities for the Drude model. With \Ref{1D} and \Ref{1dc}  we note
\be \frac{1}{\ep^{\rm D}}=\frac{\ga}{\om_p^2}\,\xi+\dots,~~~~~~\sqrt{\ep^{\rm D}}\xi R=\frac{\om_p R}{\sqrt{\ga}}\,\sqrt{\xi}+\dots\,,
\label{33D}
\ee
for $\xi\to0$. For the DC conductivity using \Ref{1dc} we have
\be \frac{1}{\ep^{\rm \DC}}=\frac{1}{\sigma}\xi+\dots,~~~~~~\sqrt{\ep^{\rm \DC}}\xi R=\sqrt{\sigma}R\,\sqrt{\xi}+\dots\,,
\label{33dc}
\ee
We see that in this approximation both models are related by the substitution $\sigma\to\om_p^2/\ga$. Therefor we can restrict ourselves to the Drude model.

\begin{figure}[h]\unitlength1cm
\hspace{0.5cm}\includegraphics[width=7cm]{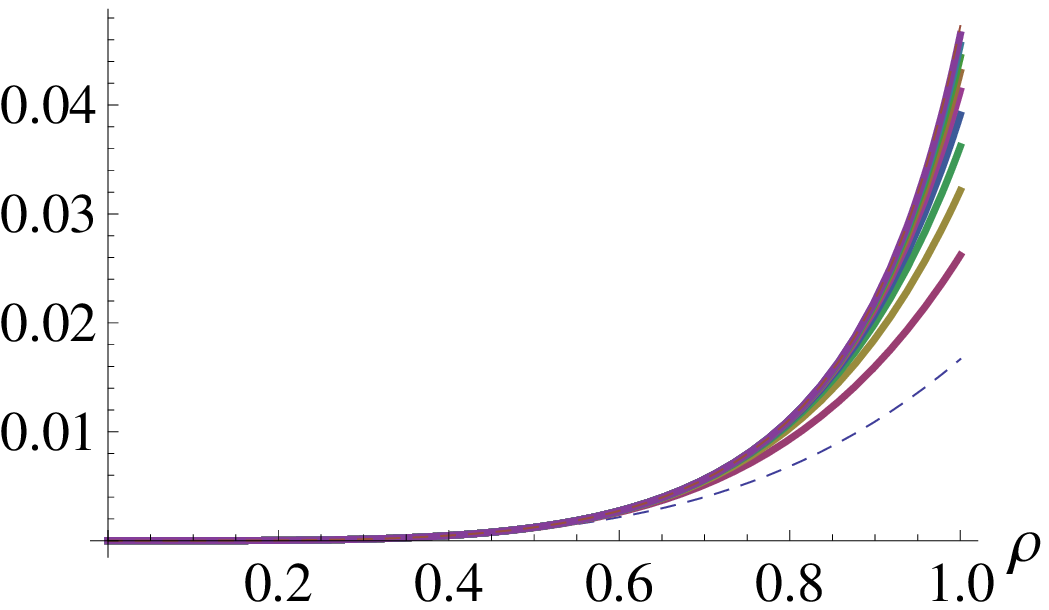}\hspace{0.5cm}\includegraphics[width=7cm]{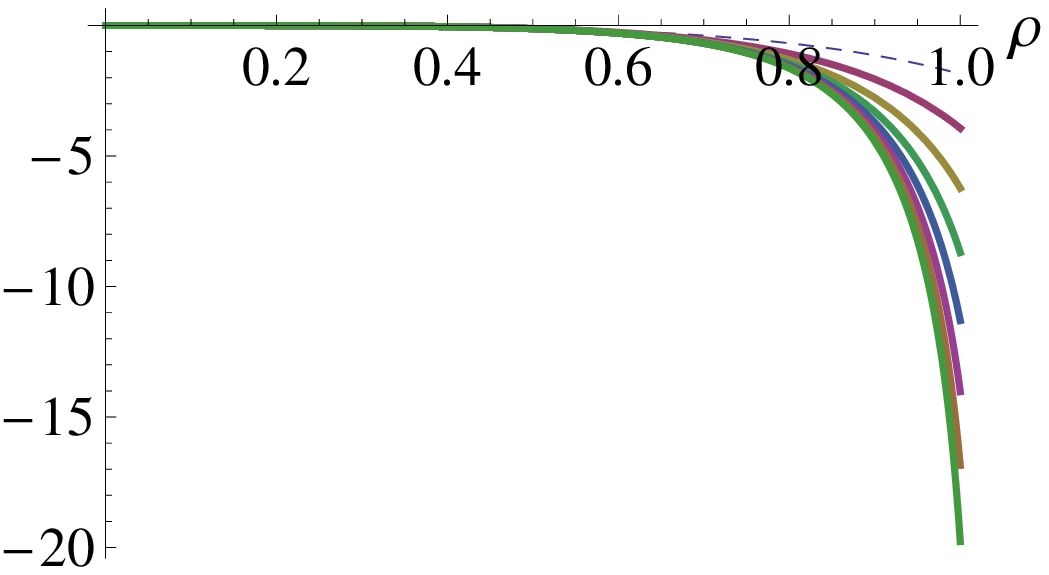}
\caption{The function $g^{\rm TE}(\eps)$ for several values of the truncation parameter, $l_m=1,\dots,11$ (left panel) and the function $g^{\rm TM}(\eps)$ for several values of the truncation parameter, $l_m=1,\dots,8$ (right panel). The dashed lines corresponds to $l_m=1$. The limit of small separation corresponds to $\eps=1$.}
\label{fig33TE}\label{fig33TM}
\end{figure}

We need to inserted \Ref{33D} into \Ref{3tTE1} and \Ref{3tTM1}. In lowest order in $\xi$ we get
\be t^{\rm TE}_l(\xi)=t_1^{\rm TE}\frac{\om_p^2R^2}{\ga}\xi+\dots
\label{33tTE}
\ee
with
\be
t_1^{\rm TE}=\frac{i_l(0)\tilde{i}^{(1)}_l -\tilde{i}_l(0) {i}^{(1)}_l}
      {k_l(0)\tilde{i}^{(1)}_l -\tilde{k}_l(0) {i}^{(1)}_l}\,,
           \label{33tTE1}\ee
for the TE mode, and
\be t^{\rm TM}_0(\xi)=t_0^{\rm TM}+t_1^{\rm TM}\frac{\ga}{\om_p^2}\xi+\dots
\label{33tTM}
\ee
with
\be
t_0^{\rm TM}=\frac{\tilde{i}_l(0) }
      {\tilde{k}_l(0)}\,,
           \label{33tTM0}\ee
and
\be
t_1^{\rm TM}=\frac{\tilde{i}_l(0) }
      {\tilde{k}_l(0)}
   \left(-1+\frac{{k}_l(0)\tilde{i}_l(0)}   {\tilde{k}_l(0){i}_l(0)}   \right)
   \,,
           \label{33tTM1}\ee
for the TM mode. We insert these expressions into the matrix $\mathbb{M}$, \Ref{3M1}, separately for both modes. For TE mode there is no zeroth order and up to the first order we find
\be \mathbb{M}^{\rm TE}= \mathbb{M}^{\rm TE}_1\ \frac{\om_p^2R^2}{\ga}\xi+\dots
\label{33MTE}\ee
with
\be  {M^{\rm TE}_1}_{l,l'}=
\frac{\sqrt{\pi}}{2}\left(\frac{\eps}{2}\right)^{l+l'+1}k_{l+l'}(0)H^{l+l'}_{l,l'}\Lambda^{l+l'}_{l,l'}
        t_1^{\rm TE}\,.
\label{33MTE1}
\ee
For the TM mode we have
\be \mathbb{M}^{\rm TM}= \mathbb{M}^{\rm TM}_0+\mathbb{M}^{\rm TM}_1\ \frac{\ga}{\om_p^2}\,\xi+\dots
\label{33MTEM}
\ee
with
\bea   {M^{\rm TM}_0}_{l,l'}&=&
        \frac{\sqrt{\pi}}{2}\left(\frac{\eps}{2}\right)^{l+l'+1}k_{l+l'}(0)
         H^{l+l'}_{l,l'}\Lambda^{l+l'}_{l,l'}  t_0^{\rm TM}\,,
         \nn\\
    {M^{\rm TM}_1}_{l,l'}&=&
        \frac{\sqrt{\pi}}{2}\left(\frac{\eps}{2}\right)^{l+l'+1}k_{l+l'}(0)
         H^{l+l'}_{l,l'}\Lambda^{l+l'}_{l,l'}  t_1^{\rm TM}\,.
\label{33MTM1}
\eea
We have to insert these expansions into the logarithm in \Ref{3F}. Expanding the logarithm we get
\be \Tr\ln\left(1-\mathbb{M}^{\rm TE}(\xi)\right)=
    -\Tr\, \mathbb{M}^{\rm TE}_1\ \frac{\om_p^2R^2}{\ga}\,\xi+\dots\,.
\label{33trE}
\ee
Finally we insert this expression into \Ref{3F} and carry out the $\xi$-integration,
\be \Delta_T\F^{\rm TE}=g^{\rm TE}(\eps)\,\frac{\om_p^2 R^2 T^2}{\ga}+\dots
\label{33DF}
\ee
with
\be g^{\rm TE}(\eps)=
    \frac{\pi}{6}\,\Tr\, \mathbb{M}^{\rm TE}_1\,,
\label{33gTE}
\ee
which is the leading order in  the low temperature expansion of the TE mode contribution to the free energy in the Drude model with fixed parameter $\ga$. The function $g^{\rm TE}(\eps)$ can be calculated numerically. For that one needs to truncate the orbital momenta $l\le l_m$. The emerging expression turns out to be converging for $\l_m\to\infty$ for all $\eps\in[0,1]$. The function $g^{\rm TE}(\eps)$ is shown in Fig. \ref{fig33TE} (left panel). In fact, Eq.\Ref{33gTE} gives a power series expansion of this function. This can be seen from \Ref{33MTE1} and the absence of $\mathbb{M}_0$ in this case.

The corresponding expression for the TM mode has to account for a nonzero $\mathbb{M}_0^{\rm TM}$ given by Eq.\Ref{33MTM1} and the expansion for $\xi\to0$ reads
\be  \Tr  \ln\left(1-\mathbb{M}^{\rm TM}(\xi)\right)
    =\Tr\ln\left(1-\mathbb{M}^{\rm TM}_0\right)
    -\Tr \left(1-\mathbb{M}^{\rm TM}_0\right)^{-1}\mathbb{M}^{\rm TM}_1\,\xi+\dots \,.
\label{33lnM}
\ee
Being inserted into \Ref{3F} only the odd term survives and after the $\xi$-integration we get
\be \Delta_T\F^{\rm TM}=g^{\rm TM}(\eps)\,\frac{\ga\, T^2}{\om_p^2}+\dots
\label{33DFM}
\ee
with
\be  g^{\rm TM}(\eps)=
    \frac{\pi}{6}\,\Tr \left(1-\mathbb{M}^{\rm TM}_0\right)^{-1}\mathbb{M}^{\rm TM}_1\,.
\label{33gTM}
\ee
This function can also be calculated numerically making a truncation as before. However, in this case the convergence is weaker. More exactly, for any fixed $\eps<1$ there is convergence for $\l_m\to\infty$, for small $\eps\gtrsim 0$ it is even very fast converging. For $\eps=1$ there is no convergence. The function $g^{\rm TE}(1)$ grows not slower than linear with $l_m$. We interpret this as a non-commutativity of the limits $T\to0$ and $\eps\to 1$. As a consequence we expect for $\eps=1$, i.e., for contact, a slower decrease with $T$ as in \Ref{33DFM}. However,  the case of contact is unphysical since in that case the vacuum energy is infinite. We have plotted the function $g^{\rm TM}(\eps)$ in Fig. \ref{fig33TM} (right panel).
%
%%%%%%%%%%%%%%%%%%%%%%%%%%%%%%%%%%%%%%%%%%%%%%%%%%%%%%%%%%%%%%%%%%%%%
\section{Conclusions}
In the preceding section we have shown that for a ball in front of a plane there is a violation of Nernst's theorem (3rd law of thermodynamics) in much the same manner as for parallel planes. For the Drude model, with a relaxation parameter $\ga$ decreasing faster than linear with the temperature $T$, the entropy at $T=0$ is given by
\be \S^{\rm D}=-\frac{1}{2\pi}\,f^{\rm D}_{\rm ball}(\eps,\om_pR)
\label{4SD}\ee
with the function $f^{\rm D}_{\rm ball}(\eps,\om_pR)$, Eq.\Ref{31f}, shown in Fig. \ref{fig3DC} (left panel). For a dielectric ball with \DC conductivity $\sigma$ also decreasing not slower than linear with $T$, the residual entropy is
\be \S^{\DC}=-\frac{1}{2\pi}\,f^{\DC}_{\rm ball}(\eps,\ep_0)
\label{4Sdc}\ee
with the function $f^{\DC}_{\rm ball}(\eps,\ep_0)$, Eq.\Ref{32f}, shown in Fig. \ref{fig3DC} (right panel). These residual entropies are in complete analogy to the planar case for which the corresponding functions are shown in Fig. \ref{fig||fD}. From here our conclusion is that the violation of the 3rd law is not related to the infinite extend of the parallel planes.  It must be mentioned that this conclusion, strictly speaking, does not apply to the Drude model since in a finite size body the relaxation parameter, which is inverse proportional to the electronic mean free path, does not vanish.

 \begin{table}[b]
\begin{tabular}{l|rcl|ccc}
    &\multicolumn{3}{c}{Drude model}&\multicolumn{3}{|c}{\DC conductivity}\\\hline
    &&&&&&\\ [-5pt]
{\bf parallel planes}&&      Eq.&Fig.&&Eq.&Fig.\\
$\ga\to0$ resp. $\sigma\to0$&&&&&&\\
vacuum Energy&TE, $\S<0$&\Ref{21Etinf}&1(left)&  TM, $\S>0$&\Ref{22Et}&  1(right)\\
$\Delta_T\F$&TE, $\S<0$&\Ref{23fD0}&2(left)&  TM, $\S>0$&\Ref{24fdc0} & 2(right)\\
$\ga$ resp $\sigma$ fixed&&&&&&\\
$\Delta_T\F$&TE, $\S<0$&\Ref{23gaf2}& &  TM, $\S>0$&\Ref{24caf2}   &\\
&           \underline{and}  TM, $\S>0$&\Ref{23gaf4}  &&&      \\
{\bf ball-plane}&&&&&&\\
$\ga\to0$ resp. $\sigma\to0$&&&&&&\\
$\Delta_T\F$&TE, $\S<0$&\Ref{31F1}&3(left)&  TM, $\S>0$&\Ref{32F1}&  3(right)\\
$\ga$ resp $\sigma$ fixed& &&&&&      \\
$\Delta_T\F$&TE, $\S<0$&\Ref{33gTE}&4(left) &   \multicolumn{3}{c}{same as Drude}    \\
&           \underline{and}  TM, $\S>0$&\Ref{33gTM} &4(right)    &\multicolumn{3}{c}{with $\frac{\om_p^2}{\ga}\leftrightarrow \sigma$}     \\

\end{tabular}
\caption{The sign of the entropy and the contributing modes for all cases considered in this paper.}
\end{table}

In order to derive \Ref{4SD} and \Ref{4Sdc} we used the representation \Ref{3F} for the temperature dependent part of the free energy as it appears from applying the Abel-Plana formula to the corresponding Matsubara sum. On that way we re-derived the violating terms for the planar case reconfirming and simplifying the original derivations, \cite{beze02-66-062112} and \cite{geye05-72-085009}. At once this allowed to extend the parameter range for which the violation occurs. We remind the two sources of entropy, $\S_0$ and $\S_1$, as defined in Eq.\Ref{2S}. $\S_0$ appears from the dependence of the vacuum energy on the temperature through the relaxation parameter $\ga(T)$ or the conductivity $\sigma(T)$ and $\S_1$ comes from $\Delta_T\F$ which is the temperature dependent part of the free energy for temperature independent $\ga$ and $\sigma$. We found that $\S_1$ has a violating term if $\ga$ or $\sigma$ decrease for $T\to0$ not slower than linear, the cases $\ga\sim T$ and $\sigma\sim T$ included. The contribution to the entropy $\S_0$ is present if the decrease is not faster than linear, i.e., $\ga\sim T^{\alpha}$ or $\sigma\sim T^{\alpha}$ with $0<\alpha\le 1$. In fact, $\S_0$ even diverges for $T\to0$. We do not discuss the question whether this has any physical relevance. Our point is simply to show what happens if inserting such parameters into the Lifshitz formula or its generalization to more complicated geometry.

We remind that the behavior of the entropy is completely different if $\ga$ or $\sigma$ have a finite limit for $T\to0$. In that case the temperature dependent part of the free energy is proportional to $T^2$ and the entropy vanishes in the limit as it should. This is well known for parallel planes and we showed  here that it holds also for a ball in front of a plane.

As observed already several times, the Casimir entropy may take negative values. In the Table 1 we collect the cases considered in this paper and show the sign of the entropy for $T$ close to zero. In addition we show the responsible mode(s)  and the relevant formulas and figures.
In all cases considered, the TE mode gives a negative contribution to the energy and the TM mode gives a positive one. In most cases only one mode contributes (the other goes with a higher power of $T$), in some cases both modes contribute to the leading behavior for $T\to0$. In that cases both signs are possible independence on the parameters involved.

%%%%%%%%%%%%%%%%%%%%%%%%%%%%%%%%%%%%%%%%%%%%%%%%%%%%%%%%%%%%%%%%%%%%%
\section*{Acknowledgement}
This work was supported by the Heisenberg-Landau program. The
authors benefited from exchange of ideas by the ESF Research Network
CASIMIR.
The authors acknowledge helpful discussions with G.Klimchitskaya and V.Mostepanenko.
I.P. acknowledges partial financial support from FRBR
grants  09-02-12417-ofi-m and 10-02-01304-a.

\bibliographystyle{unsrt}\bibliography{../../../../../Literatur/bib/papers,../../../../../Literatur/bordag}

\end{document}